\documentclass[%
 aip,
 jmp,%
 amsmath,amssymb,
 reprint,%
]{revtex4-1}

\usepackage{graphicx}
\usepackage{dcolumn}
\usepackage{bm}

\begin{document}

\preprint{AIP/123-QED}

\title{Vector Reflectometry in a Beam Waveguide}
\author{J.R. Eimer}
\altaffiliation[ ]{NASA GSRP Fellow, Goddard Space Flight Center}
\email{eimer@pha.jhu.edu}
\author{C.L. Bennett}
\affiliation{%
\mbox{Department of Physics and Astronomy, The Johns Hopkins University, Baltimore, MD}
}%
\author{D.T. Chuss} 

\author{E.J. Wollack}%
\affiliation{ 
NASA Goddard Space Flight Center, Code 665, Greenbelt, MD 20771
}%

\date{\today}

\begin{abstract}
We present a one-port calibration technique for characterization of beam waveguide components with a vector network analyzer. This technique involves using a set of known delays to separate the responses of the instrument and the device under test.
We demonstrate this technique by measuring the reflected performance of a millimeter-wave variable-delay polarization modulator.

\end{abstract}

\maketitle

\begin{quotation}
\end{quotation}

Free space reflectometry provides a non-destructive and contact-free microwave metrology technique. It is particularly well suited for measurement of the electromagnetic properties of materials and characterization of quasi-optical components in their service environments \cite{Kadaba84,Friedsam97,Gagnon03,Koers06}. However, calibration challenges exist. Errors can arise from the influence of multiple reflections, trapped modes in the system, and edge diffraction.

We present one-port data that addresses these measurement concerns using focusing elements to control the illumination of the test article, termination of the unmeasured (polarization) modes in the system, and vector calibration via a set of controlled phase delays. The method is inspired by calibration techniques employed previously \cite{Mather99,Fixsen06} and is a free-space analogue of the TRL (Thru-Reflect-Line) or more general LRL (Line-Reflect-Line) methods used for guided structures.\cite{Engen79,Marks91}  As such, extension to a multiple port calibration is straightforward. This approach also has application in the characterization of non-mating guided structures provided the electrical delays for producing appropriate sampling can be realized.

\begin{figure}
\includegraphics[width=3.25in]{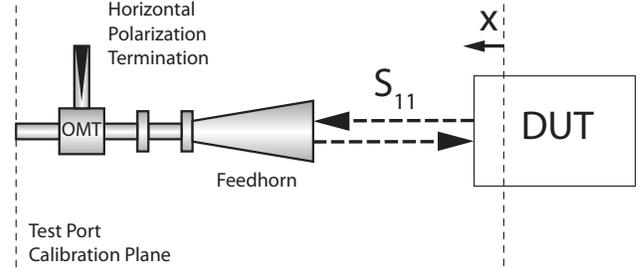}
\caption{\label{fig:concept} The system setup is shown. Dashed vertical lines indicate the nominal reference plane locations.}
\end{figure}

The non-contacting reflectometry measurement concept is illustrated in Figure~\ref{fig:concept}.  The calibration plane for the vector network analyzer (VNA) is at the interface between the single mode guide and the orthomode transducer (OMT).  The single mode from the waveguide is expanded by the feedhorn into a free space beam. The unmeasured polarization is terminated by a waveguide load and mitigates spurious resonances in the system. In this example, the signal is reflected off a device-under-test (DUT) and then coupled back into the waveguide by the same feedhorn. 

The DUT is moved through a set of distances, $x_i$, that span approximately half of a wavelength. In this way, the scattering parameter can be modeled as the complex sum of the reflection due to the instrument and that due to the DUT,
\begin{equation} 
S(x_i)=\Gamma_{inst}+e^{i2kx_i}\Gamma_{DUT}.
\end{equation}
Here $\Gamma_{inst}$ is the complex reflection coefficient from the instrument, $\Gamma_{DUT}$ is that for the DUT, $x_i$ is the translated distance for the $i$th measurement, and $k$ is the wavenumber.

It is convenient to separate these quantities into real and imaginary parts; \mbox{\( \Gamma_{inst}\equiv\Gamma_{inst}^R+i\Gamma_{inst}^I \)}, \mbox{ \( \Gamma_{DUT}\equiv\Gamma_{DUT}^R+i\Gamma_{DUT}^I \)}, and the measured data similarly can be written as \mbox{ \( \hat{S}(x_i) = \delta_i^R +i \delta_i^I \).}

If $N$ measurements are taken, a merit function can be defined,
\begin{align}
\nonumber \chi^2 &=  \sum_i^N \left(\Re[S(x_i)-\hat{S}(x_i)]\right)^2 +\left(\Im[S(x_i)-\hat{S}(x_i)]\right)^2\\
\nonumber &= \sum_i^N\left( \left(\Gamma_{inst}^R+\Gamma_{DUT}^R\cos{2kx_i} -\Gamma_{DUT}^I\sin{2kx_i}-\delta_i^R\right)^2+\right. \\ 
 &\left. \left(\Gamma_{inst}^I+\Gamma_{DUT}^I\cos{2kx_i} + \Gamma_{DUT}^R\sin{2kx_i}-\delta_i^I\right)^2 \right),
\end{align}
and minimized by solving $\nabla \chi^2 = \bf{0} $ to yield \( \bf{A \tilde{\Gamma} = D} \).
Here,
\begin{equation}
\bf{A}=\left(\begin{array}{l l l l}
N & 0 & \sum\cos{2kx_i} &-\sum\sin{2kx_i}\\
0 &N& \sum\sin{2kx_i} &  \sum\cos{2kx_i}\\
\sum\cos{2kx_i} & \sum\sin{2kx_i} & N & 0\\
-\sum\sin{2kx_i} & \sum\cos{2kx_i} & 0 &N\\\end{array} \right),
\end{equation}

\begin{equation}
\bf{\tilde{\Gamma}}=\left(\begin{array}{l l l l}
\tilde{\Gamma}_{inst}^R &\tilde{\Gamma}_{inst}^I & \tilde{\Gamma}_{DUT}^R &  \tilde{\Gamma}_{DUT}^I\end{array} \right)^T.
\end{equation}
where $\bf{\tilde{\Gamma}}$ is the best fit solution to the data, and

\begin{equation}
\bf{D}=\left(\begin{array}{l}
\sum\delta^R_i\\
\sum\delta^I_i \\
 \sum\left(\delta_i^I\sin{2kx_i}+\delta_i^R\cos{2kx_i}\right) \\
  \sum\left(\delta_i^I\cos{2kx_i}-\delta_i^R\sin{2kx_i}\right)\end{array} \right).
\end{equation}

These equations have the solution \( \bf{\tilde{\Gamma}}= \bf{A^{-1}D} \). To solve for the four components of the scattering parameter, a minimum of three delays and a reference reflection are required to span the solutions space. A similar formalism can be used for transmission measurements.

To test this technique, we measured the response of a VPM.\cite{Chuss06,Krejny08}
The VPM modulates polarization by introducing a variable phase delay between linear orthogonal polarizations. It consists of a polarizing wire grid and a mirror placed behind and parallel to the polarizer. In the system described here, the VPM wires are oriented at an angle of 45$^\circ$ with respect to the direction of the polarization of the incident radiation.  

The polarizing ``grid'' uses free-standing wires with radius $a=34$ $\mu$m and pitch $p=200$ $\mu$m. The aperture of the grid has a 50 cm diameter.\cite{Voellmer08} The face sheet of an aluminum honeycomb panel is the reflecting mirror. The grid-mirror separation and parallelism were controlled using a set of shims and verified using optical microscopy. 

\begin{figure}
\includegraphics{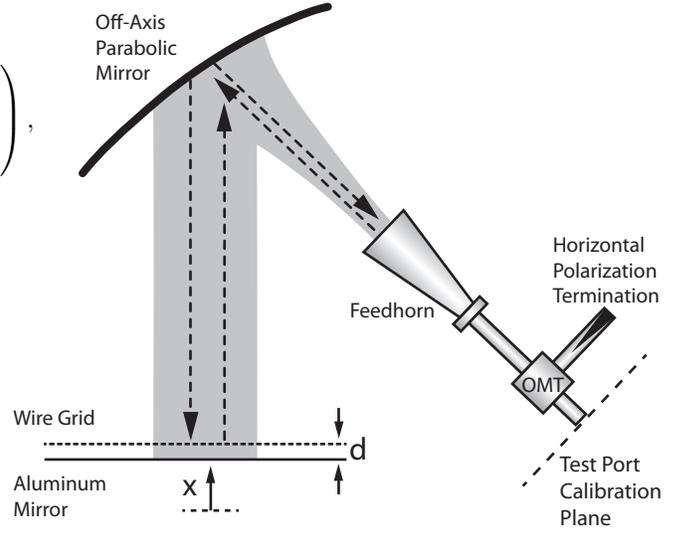}
\caption{\label{fig:setup} The test configuration is shown. The grid wires are 45$^\circ$ with respect to the nominal plane of polarization of the OMT.}
\end{figure}

The experimental setup is shown in Figure~\ref{fig:setup}. A single mode is launched from a test port of the HP 8510C network analyzer into free space using a feedhorn. An OMT\cite{Wollack02} coupled to a load is used to terminate
the unused polarization.  The radiation is collimated by a paraboloidal mirror and directed onto the center of the VPM.  It is then reflected by the VPM and refocused back into the feedhorn by the same mirror. The reflection coefficent ($S_{11}$) is measured as a function of frequency. 

The feedhorn has a gain of 25-27 dBi across the band.\cite{Barnes02} The 55 deg.\ off-axis paraboloidal mirror\cite{Wollack97} has a focal length of 50.0 cm, a diameter of ~60 cm, and is illuminated with a Gaussian beam radius of ~3.3 cm. The mirror illumination is highly apodized by the feedhorn; the edge taper for the mirror is estimated to be\mbox{ $<-40$ dB}. This configuration was dictated by available hardware and allowed many grid wires to be illuminated. 

\begin{figure}
\includegraphics[width=3.5in]{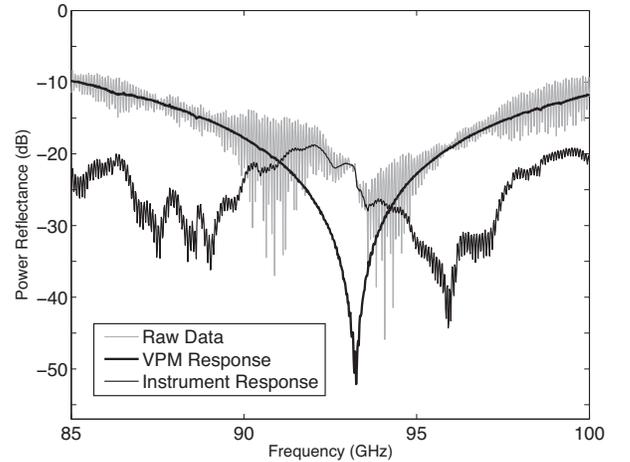}
\caption{\label{fig:response} The raw data are shown superposed on the extracted response of the VPM response and of the instrument residuals. The instrument response is dominated by the OMT.}
\end{figure}

As an example of the efficacy of the above technique, we measured the response of the VPM with the grid-mirror separation set to $d$=2,375 $\mu$m.  Using shims to adjust the distance $x_i$, we measured the response of the VPM for frequencies between 85 and 100 GHz for $x_i=\{0,200,400,600,800,1000,1200,1400,1600\}$ $\mu$m.  The results of the extraction of the device and system response are shown along with the original raw data in Figure~\ref{fig:response}.  The high spectral frequency component of the observed response in the raw data is due to the Fabry-Perot cavity formed between the OMT and its image.  Errors remaining after the calibration are dominated by the uncertainty in the $\sim \pm$25 $\mu$m determination of $x_i$.  Improved mechanical accuracy of the test setup would improve the precision of this technique.

\begin{figure}
\includegraphics[width=3.5in]{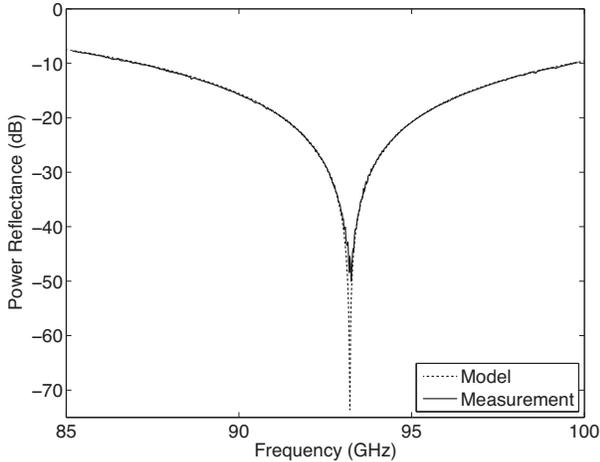}
\caption{\label{fig:model}The measured reflectance from the VPM with grid-mirror separation set to $d=$2,375 $\mu$m is normalized by the response
of the mirror alone. The model for the VPM with finite resistivity is superposed on the data. }
\end{figure}

The data in Figure~\ref{fig:response} do not take into account loss in the system.  We removed the grid from the VPM and used the aluminum plate as our reflection reference calibration standard. In Figure~\ref{fig:model} we show the calibrated VPM response corrected for system loss. The plate without the grid was measured using the same calibration introduced in this note. Residuals exist at the 0.3 dB level due to the finite directivity of the VNA modules. The data for the reflectance standard were filtered to further reduce the effect of the residual standing waves.

The VPM data are compared to a transmission line calculation for the configuration.\cite{Marcuvitz, Chuss11}  The resistance of the grid for both the capacitive (polarization perpendicular to the wires) and the inductive (polarization parallel to the wires) modes have been incorporated in the model to account for ohmic losses.

 The null in Figure~\ref{fig:model} occurs because the polarizations parallel and perpendicular to the grid wires have a relative phase of 180$^\circ$ at the frequency corresponding to the null. This is the phase difference induced by the VPM with $d=2,375$ $\mu$m. To a good approximation, the response can be modeled as a resonator with a quality factor, $Q\sim2.4$, where $Q$ is related to the measured signal, $R$, by
 \begin{equation}
 R=1-\frac{\frac{1}{(2Q)^2}}{\left(\frac{\omega-\omega_0}{\omega}\right)^2+\frac{1}{(2Q)^2}}.
 \end{equation}
 Here, $\omega$ is the angular frequency, and $\omega_0$ is the angular frequency of the null.
 The null in the reflection appears at $d/\lambda\sim0.73$ for the third cavity resonance. The deviation of this point from $d/\lambda\sim0.75$ is expected due to the phase shift of the grid circuit.\cite{Chuss11}

The null is measured at a depth of -50 dB which is higher than predicted by the simple model that is limited only by ohmic losses in the grid-mirror structure.  We find a 0.1$^\circ$ misalignment in angle of the grid relative to the axis of the OMT is sufficient to explain the measured effect. This is consistent with our confidence in the achieved rotational alignment.  Other factors contributing to the observed finite reflectance at the null include the limited precision of the measurement of $x_i$ and the limitations on dynamic range and readout resolution of the HP 8510C.
In applying this technique to the characterization of high \emph{Q} systems, changes in optical loss as $x_i$ is varied would be of greater importance. For systems with high reflectance, this technique could be improved by adopting a bilinear form for the reflectance.

In this paper we have developed and demonstrated a vector calibration technique that improves beam wave\-guide metrology. For single-moded measurements, we have improved the precision over currently-employed techniques.  The method described here can be extended to 2-port systems.

We would like to thank G. Voellmer, H. Hui, and L. Zeng for experimental setup, L. Page for access to optical components, and D.J. Fixsen for helpful discussions.

%

\end{document}